\documentstyle[prl,aps,preprint,epsf]{revtex}
\tightenlines
\begin{document}
\draft
\title{
	The fully frustrated XY model with next nearest neighbor interaction
}
\author{
	Giancarlo Franzese$^{(1,2,3)}$ \thanks{Present address: Center
for Polymer Studies, 
Boston University, 590 Commonwealth Avenue, Boston, MA 02215}, 
	Vittorio Cataudella$^{(1,2)}$,  
	S.E. Korshunov$^{(4)}$, and Rosario Fazio$^{(1,5)}$
}
\address{
	$^{(1)}$Istituto Nazionale per la Fisica della Materia (INFM),
	Unit\`a di Napoli e Catania\\
	$^{(2)}$Dip. di Scienze Fisiche, Universit\'a ``Federico II",
	Mostra d'Oltremare, Pad. 19, 80125 Napoli - Italy\\
	$^{(3)}$Dip. di Fisica ``E.Amaldi'', Universit\`a Roma Tre,
	via della Vasca Navale 84, 00146 Roma - Italy\\
	$^{(4)}$L.D. Landau Institute for Theoretical Physics, Kosygina 2,
        117940 Moscow, Russia\\	
	$^{(5)}$Dip. di Metodologie Fisiche e Chimiche (DMCFI),
	Universit\`a di Catania, viale A. Doria 6, 95125 Catania - Italy
}
\date{\today}
\maketitle
\begin{abstract}
We introduce a fully frustrated XY model with nearest neighbor
(nn) and next nearest neighbor (nnn) couplings which can be realized in
Josephson junction arrays. We study the phase diagram for $0\leq x \leq 1$ ($x$
is the ratio between nnn and nn couplings). When $x < 1/\sqrt{2}$
an Ising and a Berezinskii-Kosterlitz-Thouless  transitions are present. 
Both critical temperatures decrease with increasing $x$. For $x > 1/\sqrt{2}$
the array undergoes a sequence of two transitions.
On raising the temperature first the two sublattices decouple from each other and 
then, at higher temperatures, each sublattice becomes disorderd.
\end{abstract}


\narrowtext
A variety of two dimensional systems undergo a phase transition without
a rigorous symmetry breaking, the Berezinskii-Kosterlitz-Thouless (BKT)
transition~\cite{BKT}. The transition is driven by the thermally excited 
vortices which form a two dimensional Coulomb gas~\cite{minnhagen}.
Josephson junction arrays are experimental realization of the two-dimensional
XY model where the array's parameters can be modified in a controlled way.
In the last decade there has been a great amount of work on the various
aspects of the  BKT transition in Josephson arrays~\cite{proceeds}.
Experimental studies are based on electrical resistance~\cite{exp-resistance},
two-coil inductance~\cite{exp-inductance}, and SQUID~\cite{shaw} measurements.

A magnetic field applied perpendicularly to the array leads to
frustration~\cite{FRUSTRATION,halsey}. If the flux piercing the elementary
plaquette is half of the flux quantum $\Phi _0=hc/2e$, the system is  called
fully frustrated (FF) and undergoes two phase transitions
related to the $Z_2$ and $U(1)$ symmetries. The existence of the two critical
temperatures $T_c^{Z_2}$ and $T_c^{U(1)}$ respectively has been extensively
investigated both by analytical methods~\cite{halsey,night} and Monte Carlo (MC)
simulations~\cite{teitel,granato1,jose,olssonf1,grest}.
The complete scenario is not fully understood yet. There are numerical evidences
either supporting the existence of two very close critical temperatures
($T_c^{Z_2}>T_c^{U(1)}$) with critical behavior typical of Ising and  BKT
transitions respectively~\cite{olssonf1,grest,olssonf2} or the existence of a
single transition with novel critical behavior~\cite{granato1}.

Untill recently only the Josephson arrays with nearest neighbor (nn)
couplings were studied. Recent theoretical~\cite{chandra}
and experimental~\cite{shea} works on infinite range array opened a new field
of investigation in these systems. In this Letter we study the properties of a
two dimensional FF Josephson array with both nn and
next-to-nearest neighbor (nnn)
couplings~\cite{competing}. Proximity-junction arrays may be good candidates to
experimentally probe the effects discussed in this work. They consist of
superconducting islands in good electric contact with a metallic substrate.
Due to the proximity effect there is a leakage of Cooper pairs into
the normal substrate which extends over a temperature-dependent coherence
length $\xi _N$ ($\xi _N=\hbar v_F/k_BT$ or $\xi _N=\sqrt{\hbar D/2 \pi k_BT}$
for the ballistic and for the diffusive case respectively, $v_F$ is the Fermi
velocity, $D$ the diffusion constant and $T$ the temperature). When $\xi _N$
becomes comparable with the lattice constant of the array
the nnn coupling becomes comparable with the nn coupling.
Since $\xi _N$ is strongly
dependent on the temperature, the nnn Josephson coupling may be observed cooling
down the sample. The main results of this Letter are summarized in the phase
diagram of Fig.~\ref{fig5}.

The system is defined by the Hamiltonian
\begin{eqnarray}
	H = -\sum_{<<i,j>>}J_{ij}\cos ( \theta _i-\theta _j-A_{ij})
\label{model}
\end{eqnarray}
where $J_{ij}=J>0$ for nn and $J_{ij}=xJ$ for nnn ($x\geq 0$), the symbol 
$<<\ldots>>$ refers to the sum over nn and nnn. For later convenience we introduce
the gauge invariant phase difference  $\phi_{ij} = \theta _i-\theta _j-A_{ij}$.
The variables $\theta_i $ are the phases of the superconducting order
parameter of the i-th island. The magnetic field enters through
$A_{ij}=(2\pi/\Phi_0)\int_i^j{\bf A} {\bf \cdot }d{\bf l}$ (${\bf A}$ is
the vector potential). The relevant parameter which describes the magnetic
frustration is $f=\sum A_{ij}/(2\pi)$, where the summation runs over the
perimeter of the elementary plaquette. We study the case $f=1/2$ on a square
lattice.

\underline{Ground States} -
The model of Eq.(~\ref{model}) combines the characteristics of
both the FF and unfrustrated XY models. While the elementary square plaquet\-te
is FF, the square plaquet\-tes formed  by nnn couplings are not frustrated.
For $x<x_0\equiv 1/\sqrt{2}$ we find that the ground state of this system is
exactly the same as in the FF model without nnn couplings and is characterized by
$\phi_{ij} = \pm  \pi  /4$ for  all pairs of nn sites.
This state combines continuous [$U(1)$] and discrete  ($Z_2$) degeneracies. 
Its energy per site $E=-\sqrt{2}J$ is independent of nnn coupling $xJ$.
Moreover if one considers a straight domain wall separating the two ground
states with opposite orientations of chiralities it turns out that
neither the form of such state nor its energy change with addition of
nnn interactions.

For $x>x_0$ the ground state is the same as in the absence of nn coupling
when the system splits into two unfrustrated $XY$-models.
The relative phase shift between the two sublattices can be
arbitrary. The energy of this state $E=-2xJ$ depends only on nnn
coupling $xJ$ and does not depend on nn coupling.

\begin{figure}
\begin{center}
\mbox{\epsfxsize=12cm \epsfysize=7cm \epsffile{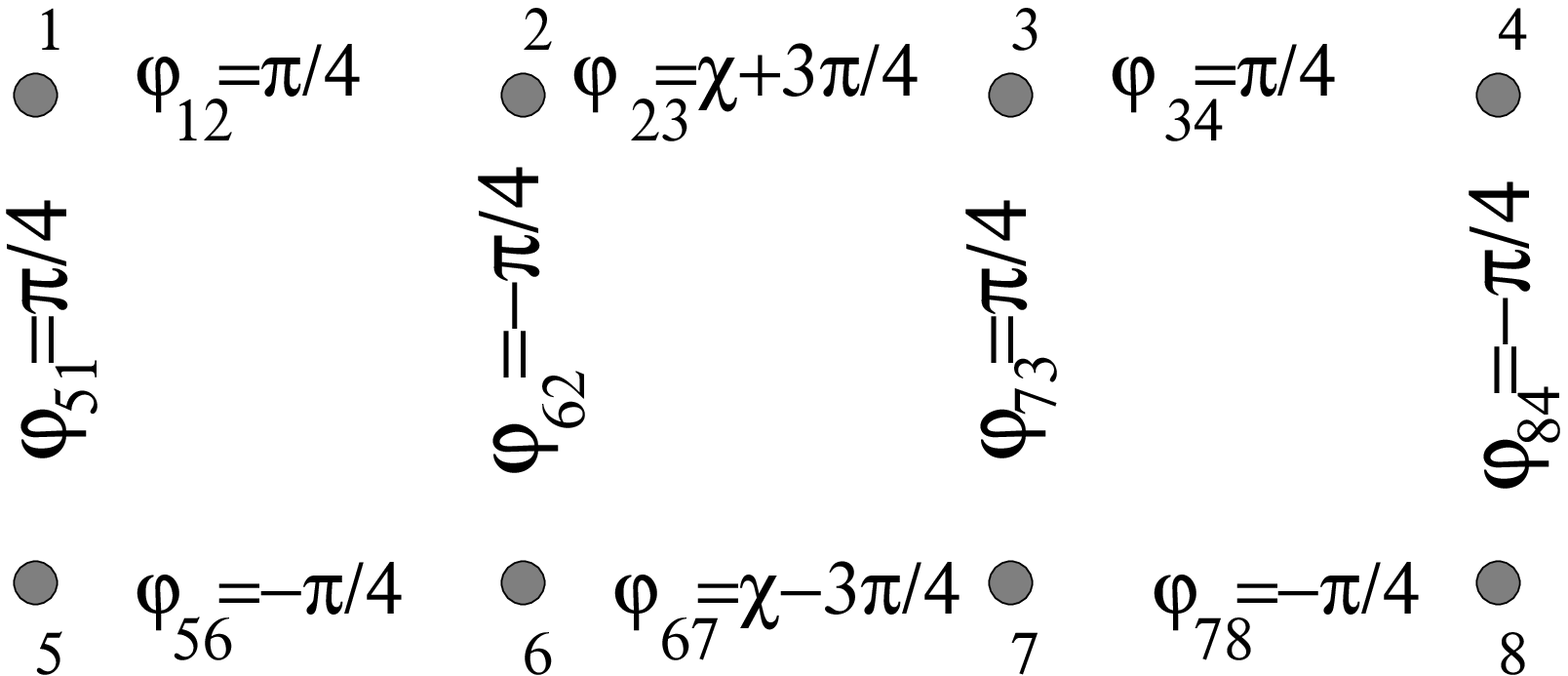}}
\caption{The family of the additional (intermediate) ground
	states at the degeneracy point $x=1/\protect\sqrt{2}$ is characterized by
	the angle $\chi$. The value $\chi=\pi$
	corresponds to low-$x$ ground state while  $\chi=0$ corresponds to high-$x$
	ground state with a particular relative  phase shift between
	the two sublattices.
	The angle reported in the figure is the gauge invariant phase difference
	between two neighboring sites.
}
\label{fig1}
\end{center}
\end{figure}

At the special point $x=x_0$ the energies of both the above ground states
coincide. Morover they can be transformed into each other by a continuous
transformation without increasing the energy  of the system. Therefore
the manifold of the ground states also includes an additional set of
eight-sublattice "intermediate" states which can be parametrised
by a rotation angle $\chi$ ($\chi=\pi$ corresponding to low-$x$ ground state
and $\chi=0$ corresponding to high-$x$ ground state with a particular
relative phase shift between the sublattices) as it is shown in
Fig.~\ref{fig1}.

\underline{Phase Diagram} - We studied the finite temperature behaviour
of the model by means of  the (low temperature) spin-wave free energy analysis
and by Monte Carlo simulations.

 For $x< x_0$  neither the spectrum of spin waves (in the long wavelength limit)
nor the domain wall energy depend on $x$. Therefore we  can expect  only a weak 
dependence of $T^{Z_2}_c$ and $T^{U(1)}_c$ on $x$, due to the change of the 
effective interaction between the different types of fluctuations.

For $x>x_0$ the system (at finite temperatures) turnes out to be equivalent to
two coupled XY models, the effective forth-order coupling between the two sublattices
provided by the free energy of spin waves. Although this coupling is weak (always
much smaller than the temperature) at low temperatures it is relevant and imposes
the presence of a transition at $T=T_{D}$. This transition separates the phases with
coupled and decoupled sublattices. In the low temperature phase, where the two
sublattices are locked, the spin-wave contribution to the free energy imposes
a relative phase shift of $\pm\pi/4$ (or equivalently $\pm 3\pi/4$)
between the two sublattices. At $ T>T_D$
a second phase transition of the BKT type takes place in each of the decoupled
sublattices. For $x\gg x_0$ the temperature of this transition depends
only on nnn coupling and it is proportional to $x$.

The spin wave spectrum remains rigid down to $x=x_0$. This indicates (and it
is confirmed by the MC simulations) that the critical temperature of
this BKT transition remain finite when $x\rightarrow x_0^+$. Below this
temperature there is a transition between the FF low-$x$ and the
unfrustrated high-$x$ phases. We evaluated  numerically the spin wave free energy
of the intermediated ground state as a function of $\chi$.
This dependence is described by a convex function implying a first order
phase trasition line separating the low-$x$ and high-$x$  quasi-ordered
phases.

The critical properties were investigated evaluating the helicity modulus $\Gamma$
and the staggered chiral magnetization $M$ by means of  standard MC simulations for
different $x$. The order parameter $M$, which controls the Ising-like transition,
is defined as
\begin{equation}
	M=\frac 1{L^2}\left| \sum_{i}(-1)^{i_x+i_y}m_{i}\right|   \label{m}
\end{equation}
where $\vec{r}_i/a=(i_x,i_y)$ is the position vector (in unit of lattice step
$a$) of the site $i$ and 
$
	m_{i}=\frac 1{\sqrt{8}}(\sin\phi _{i,j_1}+
		\sin\phi _{j_1,j_2}
		+\sin\phi _{j_2,j_3}+
		\sin\phi _{j_3,i})
$
is the chirality of the plaquette with center in $(i_x+1/2,i_y+1/2)$ and
with site indexes $i$,$j_1$,$j_2$,$j_3$ (in clockwise order).

\begin{figure}
\begin{center}
\mbox{\epsfxsize=10cm \epsfysize=10cm \epsffile{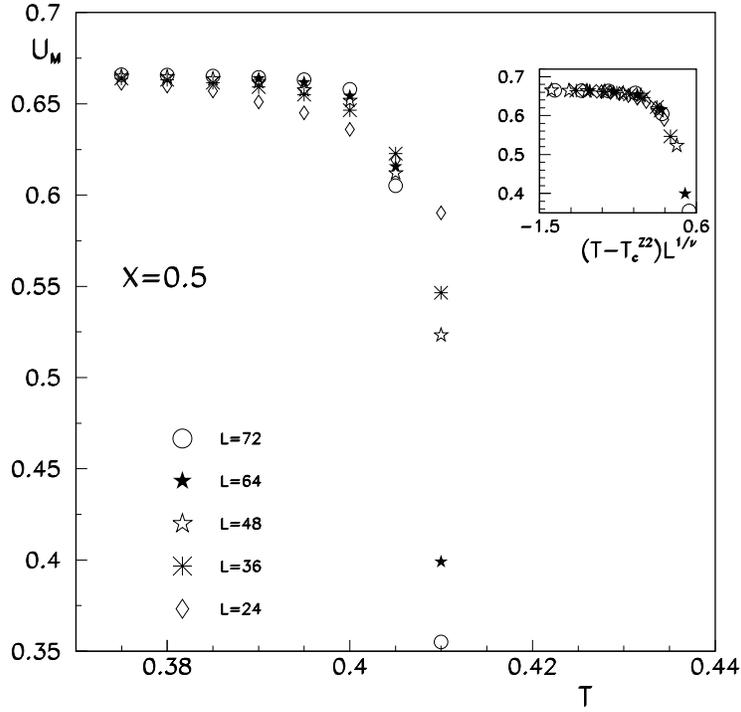}}
\caption{The  Binder's parameter $U_M$ vs. $T$ for several lattice sizes
	$L$ at $x=0.5$. Errors are smaller than the symbols size.
	Excluding the data of smallest size
	($L=24$) one can estimate $0.400< k_BT_c^{Z_2}/J <
	0.406$. Inset: Collapse of the data  (excluding the $L=24$ points).
	The scaling parameters are $k_BT_c^{Z_2}/J=0.403\pm 0.003$ and
	$1/\nu=1.0\pm0.1$.
}
\label{fig2}
\end{center}
\end{figure}

\begin{figure}
\begin{center}
\mbox{\epsfxsize=10cm \epsfysize=12cm \epsffile{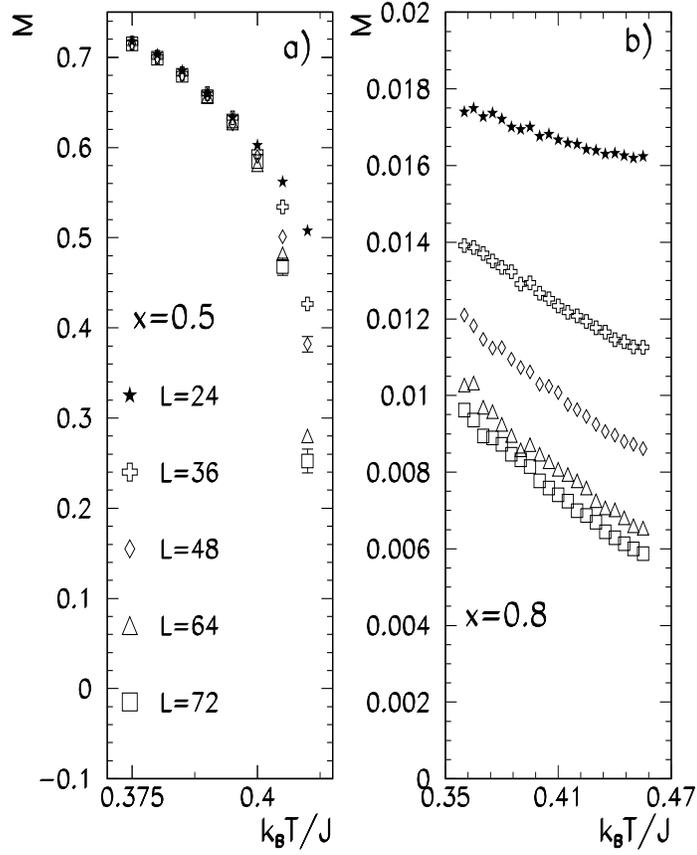}}
\caption{The staggered chiral magnetization $M$ vs. $T$
	for $x=0.5$ a) and $x=0.8$ b). For large $L$ and low $T$, $M$ goes
	to a non-zero value for $x=0.5$ and vanishes for $x=0.8$.
	The errors are smaller than the symbols size. The symbols for different
	lattices sizes are the same in a) and b).
}
\label{fig3}
\end{center}
\end{figure}

The helicity modulus $\Gamma= \partial^2 {\cal F}/\partial \delta^2$, used to
signal the existence of a  BKT transition, is defined through
the increase of the free energy ${\cal F}$ due to a phase twist
$\delta$ imposed in one direction~\cite{Ohta}.
In order to obtain a precise determination of $T_c^{Z_2}$ and of
the critical exponent $\nu$ associated to the divergence of the correlation
length we have calculated the Binder's cumulant~\cite{binder} of the staggered
chiral magnetization $M$
\begin{equation}
	U_M=1-\frac{\langle M^4\rangle }{3\langle M^2\rangle ^2}.
\label{Ubinder}
\end{equation}
Since  $U(T_c^{Z_2},L)$ does not depend on lattice size $L$ for large systems,
$T_c^{Z_2}$ can be identified without making any assumption on the critical
exponents. Once a satisfactory estimation of $T_c^{Z_2}$ is obtained the critical
exponent $\nu $ is estimated through a data collapsing with $\nu$ left as the only
free parameter. Estimation of $U_M$ have been obtained averaging,
at least, $10^7\cdot L^2$ MC configurations by using a standard Metropolis algorithm.
The largest lattice studied is $L=72$. 
The result for $x=0.5$ are shown in Figs.\ref{fig2},\ref{fig3}a.  
We estimate $k_BT_c^{Z_2}/J=0.403\pm 0.003$.
The data collapsing, shown in the inset of Fig.\ref{fig2}, gives an estimate of
$1/\nu=1.0\pm 0.1$. The critical temperature $T_c^{Z_2}$ decreases with increasing 
$x$ for $x<x_0$. For $x\geq x_0$, there is no sign of a Ising-like transition (see
Fig.\ref{fig3}b). 

\begin{figure}
\begin{center}
\mbox{\epsfxsize=10cm \epsfysize=12cm \epsffile{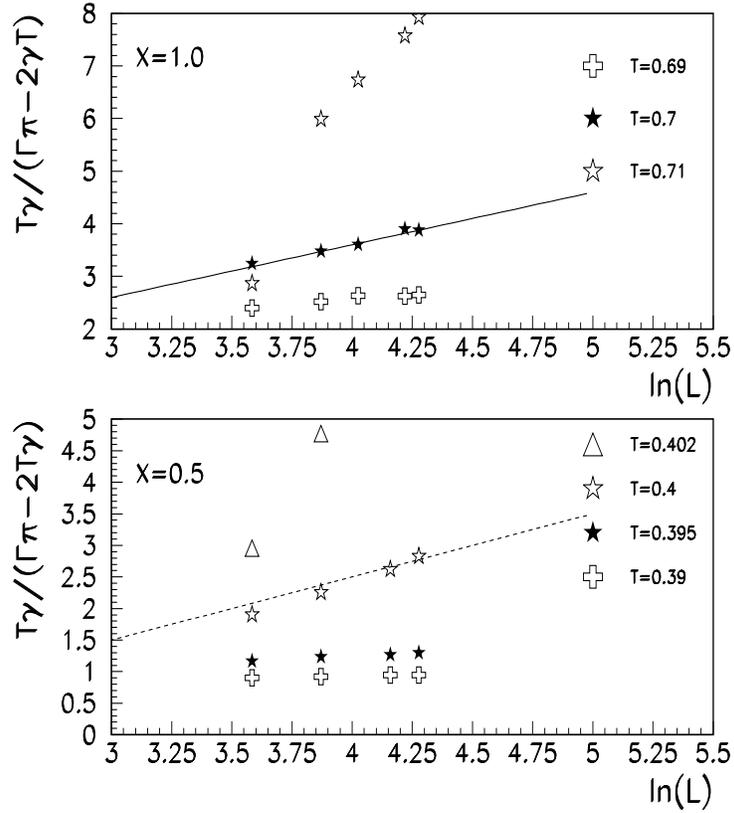}}
\caption{The size dependence of the helicity modulus $\Gamma$
	for $x=0.5$ and $x=1$ at several temperatures $T$.
	The errors are smaller than the symbols size.
	The form of the plot [$\gamma T/(\pi\Gamma-2\gamma T)$ vs. $\ln
	L$] is chosen in such a way that the scaling behaviour predicted by
	Eq.~(\ref{ansatz}) should correspond to a straight line with the slope
	equal to 1 (as shown by the over-imposed lines).
	The estimates for the critical temperature $T_c^{U(1)}$ are
	$0.4 \pm 0.002$ and  $0.7 \pm 0.005$ respectively. The error 
	on the estimates is due to the used temperature mesh.
}
\label{fig4}
\end{center}
\end{figure}

\begin{figure}
\begin{center}
\mbox{\epsfxsize=12cm \epsfysize=12cm \epsffile{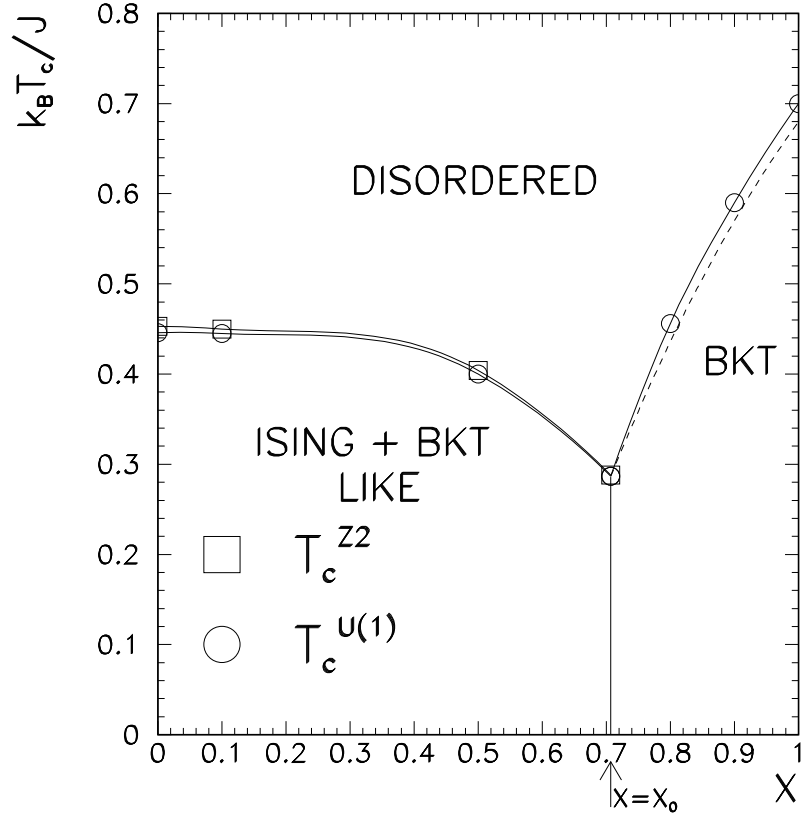}}
\caption{The phase diagram for the XY model with nn+nnn
	interactions. Squares refer to $T_c^{Z_2}$, circles to $T_c^{U(1)}$;
	$x$ is the ratio between nnn and nn couplings; $J$ is the nn
	coupling. The errors are smaller than the symbols size (see text for 
	details). The dotted line  shows the qualitative behaviour of the
	transition associated to the decoupling of the two nnn sublattices.
}
\label{fig5}
\end{center}
\end{figure}

Following the procedure proposed in Refs.~\cite{hm,olsson_hm},
the critical temperature $T_c^{U(1)}$ is estimated by using the following 
{\em ansatz} for the size dependence of $\Gamma $
\begin{equation}
	\frac{\pi \Gamma }{2T_c}=
	\gamma\left[1+\frac {1}{2(\ln L-\ln l_0)}\right]
\label{scaling}
\label{ansatz}
\end{equation}
where $l_0$ is a fit parameter.
This critical scaling is based on the mapping between a neutral Coulomb gas
and a  XY model. Therefore Eq.(\ref{scaling}) can be used both as a
test for the existence of a BKT transition and for a precise evaluation of
the critical temperature. A very good scaling was obtained with $\gamma=1$ 
(the ordinary BKT transition) in the low-$x$ phase and $\gamma=2$ (corresponding to
BKT transition on each  of the two sublattices  with lattice constant
$\sqrt{2}$) in the high-$x$ phase.
In Fig.\ref{fig4} we show this analysis for the cases $x=0.5$
and $x=1$. $ T_c^{U(1)}(x)$, as well as $T_c^{Z_2}(x)$, decreases with 
increasing $x$ up to $x \sim x_0$. Our results cannot  discriminate between the 
$T_c^{Z_2}=T_c^{U(1)}$ and the $T_c^{Z_2} > T_c^{U(1)}$ hypothesis since the two 
temperatures are compatible within the numerical precision (the mean value of 
$T_c^{Z_2}(x)$ remains always above the corresponding mean value of $T_c^{U(1)}(x)$).
For $x\geq 0.8$, instead, $T_c^{U(1)}(x)$ increases, quickly tending towards 
the value expected for $x\rightarrow \infty$ i.e. $k_B T_c^{U(1)}/(xJ)=0.89 $.

We finally discuss the transition related to the decoupling of sublattices in
the high-$x$ phase. The order parameter $S=\sum_{j,\alpha}s_{j,j+e_\alpha} $
can be defined on the bonds of the lattice and can be chosen in the following
gauge-invariant form:
\begin{eqnarray}
	s_{j,j+e_x} &=&(-1)^{j_x}\exp\left[i(-1)^{j_x+j_y}\phi_{j,j+e_x}\right]
	\nonumber \\
	s_{j,j+e_y}&=&i(-1)^{j_y}\exp\left[i(-1)^{j_x+j_y}\phi_{j,j+e_y}\right]
	\nonumber
\end{eqnarray}
This form of the order parameter is chosen in such way that for any
of the high-$x$ ground states the value of $s$ will be the same
for all the bonds. At low temperatures $S$ manifests a true long-range order.
The MC simulations confirm the $\pi/4$ relative  phase
shift anticipated by the spin-wave analysis.
By increasing the temperature the long-range order in $S$ can be expected
to disappear as a separate phase transition whereas the unbinding of vortex
pairs in each of the sublattices has to occur at still higher temperatures.
We performed MC simulations to evaluate the transition temperature $T_D$.
The results of this computation (not reported here)  show that $T_D$ is very close,
but lower, than the transition to the disordered phase $T_c^{U(1)}(x)$.
More extensive simulations are needed to determine the critical behaviour of
decoupling transition. The dotted line in Fig.\ref{fig5} shows the qualitative
behaviour of the decoupling transition as a function of $x$.

In conclusion we have introduced a frustrated XY model with nnn interaction.
The model can be experimentally realized in Josephson junction arrays
in a transverse magnetic field. Signatures of nnn Josephson couplings might
already have been seen in specially designed setups~\cite{communication}.
The analysis presented here leads to the phase diagram shown in Fig.\ref{fig5}.
For $ 0<x<x_0$  the critical temperatures  $T_c^{Z_2}$ and
$T_c^{U(1)}(x)$ decrease with
increasing $x$. For $x > x_0$, there is no sign of a Ising-like transition
and the system  behaves like the unfrustrated XY model. At $x\approx
x_0$  in the low $T$ region there is a first order phase transition between the 
low and high-$x$ phases. Finally for $x > x_0$ the array undergoes a sequence 
of two transitions. On raising the temperature, first the two sublattices become 
decoupled and then, at higher temperatures, each sublattice becomes disordered.

We thank H.~Courtois and B.~Pannetier for fruitful discussions.
We acknowledge the financial support of INFM (PRA-QTMD) and the
European Community (Contract FMRX-CT-97-0143).
G.F. is grateful to A.~Mastellone and J.~Siewert for
hospitality in Catania.



\newpage

\newpage

\newpage

\newpage

\end{document}